# On Mobile Cloud for Smart City Applications

Manfred Sneps-Sneppe
Ventspils International Radioastronomy Centre
Ventspils University College
Ventspils, Latvia
manfreds.sneps@gmail.com

Dmitry Namiot
Faculty of Computational Mathematics and Cybernetics
Lomonosov Moscow State University
Moscow, Russia
dnamiot@gmail.com

*Abstract*—This paper is devoted to mobile cloud services in Smart City projects. As per mobile cloud computing paradigm, the data processing and storage are moved from the mobile device to a cloud. In the same time, Smart City services typically contain a set of applications with data sharing options. Most of the services in Smart Cities are actually mashups combined data from several sources. This means that access to all available data is vital to the services. And the mobile cloud is vital because the mobile terminals are one of the main sources for data gathering. In our work, we discuss criteria for selecting mobile cloud services.

## I. INTRODUCTION

In this article, we would like to focus on selecting solutions for mobile cloud services in the smart city. We discuss criteria for selection and rationale for architectural decisions in the pilot project for a mobile operator in Moscow.

As per [1], mobile cloud computing is a paradigm for mobile applications whereby the data processing and storage are moved from the mobile device to centralized computing platforms located in clouds. These centralized applications are then accessed over the wireless connection based on a thin native client or web browser the mobile devices. So, it is an infrastructure where both the data storage and data processing happen outside of the mobile device.

As per definition from British Standard Institute [2], Smart City is an effective integration of physical, digital and human systems in the built environment to deliver a sustainable, prosperous and inclusive future for its citizens.

A key feature of smart cities is the ability of the component systems to interoperate. The above-mentioned PAS-182 [2] defines a concept model and gives guidance to decision makers on applying it to promote interoperability for data created, used, and maintained by a city across all sectors.

So, it is a vital issue for services in Smart Cities to share data. Even more, we can conclude that most of the services for Smart Cities are mashups and collect (proceed) data from many sources [3]. In the above-mentioned paper, we show that most services could be classified via Data Program Interfaces (DPI), rather than via Application Program Interfaces (API).

As per BSI specification, data is a resource that can transform the capability of a city, enabling the development of systems and services, and supporting informed decisions. And the centralized store (cloud at the whole is centralized) is the convenient way for access to different data from mashups. In general, Smart City consists of organizations across all sectors, facilitated by the sharing of data, based on a common framework of its meaning, and consistent use of identifiers and classifications [2].

Why should it be a mobile cloud? Mobile devices play an important role in gathering data in Smart Services. It is so-called crowd-sensing (or mobile crowd-sensing) [4].

There is so-called crowd-sourcing as a form of cooperation of a group of users (crowd), where all single users are performing small subtasks of a bigger task. It lets handle complex problems with many co-working users. Crowd-sensing is a subtype of crowd-sourcing where the actually outsourced job is a complex sensing task [5].

Crowd-sensing could be used in Smart Cities alone and alongside with sensor networks [6]. It is an additional technology which involves moving sensors on mobile devices. In Smart Cities, many crowd-sensing applications target such areas as noise and pollution measurements, urban transportation systems, tracking of public buses and trams, etc. The typical tasks for mobile tracking are circled about various forms of monitoring [7].

Of course, we need to save data from mobile sensors. And data should be available for mashups too. It means that mobile cloud should be an important part of Smart City platform. It could bring the following benefits for Smart Cities [2]:

- reduced cost as the need to recollect and verify data is removed;
- integrated city systems and data-driven services;
- a common understanding of the needs of communities;
- shared objectives, collaboratively developed and evidenced using data;
- engaged and enabled citizens and communities;
- transparency in decision-making models;

- consequently improved quality of life for citizens.

The rest of the paper is organized as follows. In Section II, we discuss related works. In Section III, we discuss criteria for selecting mobile cloud solutions in Smart Cities. In Section IV, we present our architecture for mobile cloud in Smart Cities.

## II. RELATED WORKS

With the original idea to eliminate the constraints of weakness in computing power in mobile devices, mobile cloud computing is an attractive topic in scientific research as well as in practical implementations.

Mobile cloud computing has various service models [8]. In the upper layers, we have the following paradigms: Data centers layers, Infrastructure as a Service (IaaS), Platform as a Service (PaaS), and Software as a Service (SaaS) (Figure 1).

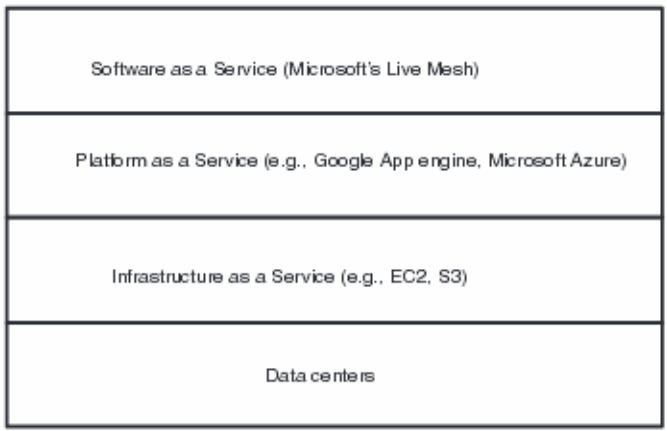

Fig. 1 Cloud computing paradigm

Data centers layer provides the hardware facility and infrastructure for clouds. IaaS is built on top of the data center layer and enables the provision of storage, hardware, servers, and networking components. The canonical examples of IaaS are Amazon Elastic Cloud Computing and Simple Storage Service (S3) [9]. PaaS offers an advanced integrated environment for building, testing, and deploying custom applications. The examples of PaaS are Amazon Map Reduce/Simple Storage Service, Google App Engine or Microsoft Azure [10]. SaaS supports a software distribution with specific requirements. The canonical examples are Salesforce or Google Apps [11].

For smart cities at the present time, we have a very agile structure of services, which is continuously added new services and removed the old ones. In addition, the idea of smart cities implies an increase in the number of developers from many different areas (we need mashups). It means that PaaS should be the most suitable layer for Smart Cities backends unless we will get a stable set of services and move them to SaaS layer.

What is the main difference of mobile clouds from the general model? We have to have some level of support for mobile devices (mobile operations systems). So, for the last years, we can see the emergence of a new technology (and new acronym), Mobile Backend as a Service (MBaaS) [12].

MBaaS (backend as a service - BaaS), is a model for providing the web and mobile app developers with a way to link their applications to backend cloud storage [13]. MBaaS exposes APIs and custom software development kits (SDKs) for mobile developers and also provides features such as user management, push notifications, and integration with social networking services. Usually, MBaaS providers offer a different set of backend tools and resources. As the common services provided by the majority of providers, we can mention file storage, file sharing, push notifications, location services, chat and messaging, integration with social networks such as Facebook and Twitter, usage analysis tools [14]. MBaaS are gaining mainstream traction with enterprise consumers being vendor-agnostic and suitable for novice developers.

For example, MBaaS Convertigo [15], supports a wide set of features. Developers can connect to enterprise data using a wide range of connectors such as SQL or Web Services. MBaaS supports cross-platform development – desktop and mobile apps on multiple devices (iOS, Android), as well as server-side business logic. Convertigo supports push notifications and test driven development, integrated version control (GIT, SVN), etc.

The common features for MBaaS include also support for programming device features (e.g., plugins, APIs) such as cameras or sensors, support for development environment (e.g., integrated version control or GIT), multiple operational systems support, cloud deployment, testing support and encrypted transactions. An important feature for all mobile backends is activity monitoring. It lets monitor system activities such as connected devices, server's request or response's time and logging. This monitor should also support a search for activity logs with rich tracking and filtering options. Of course, MBaaS should support user authentication (e.g., LDAP, Facebook Connect), mobile applications management, visual development and provide task scheduler. The scheduler is mandatory at least for push notifications planning [16].

As per [17], MBaaS offerings sit squarely between the existing platform-as-a-service vendors and the full end-to-end solution space occupied by mobile enterprise/consumer application platforms (Figure 2).

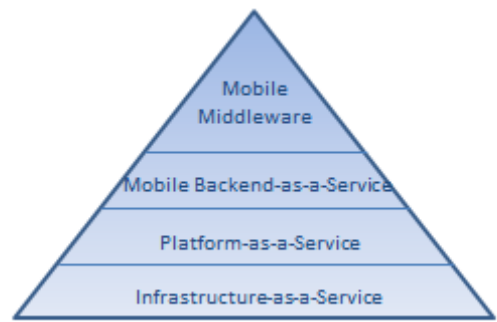

Fig. 2 MBaaS triangle [17]

As a typical example here we can mention FI-WARE [18] mobile cloud. It is based on OpenStack functionality. We are mentioning FI-WARE because it is a "standard" offer for Smart City projects, supported by European Commission.

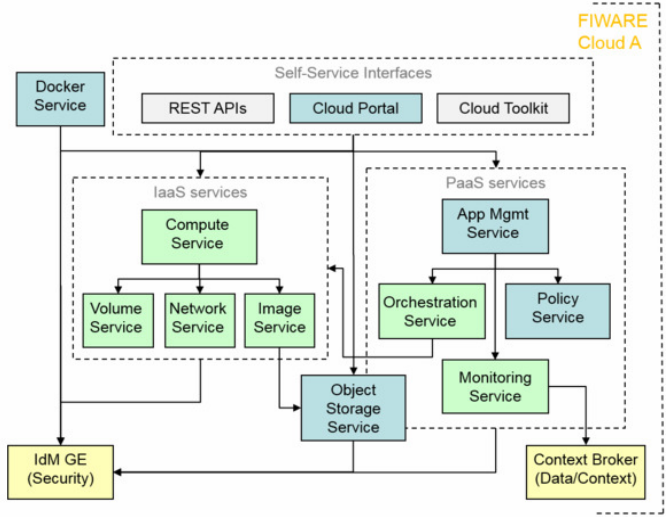

Figure 3. FI-WARE mobile cloud [19]

We've mentioned already, that by our opinion FI-WARE is over-engineered and unnecessary complexity [3]. In Figure 4, we illustrate the MBaaS platform from EMC [20]

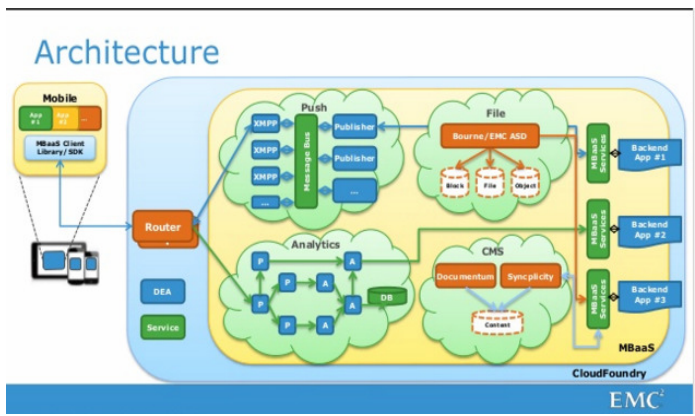

Figure 4. EMC MBaaS [20]

It is more "service" oriented and directly enlists proposed services as push notifications, analytics, file store, etc. In the same time, FI-WARE offers a generic approach and leaves details to so-called enablers. Actually, it is a very important point of view. Excessive generalization loses more practical solutions.

The Cloud Standards Customer Council (CSCC) [21] has published the Customer Cloud Architecture for Mobile whitepaper. This paper describes vendor-neutral best practices for hosting the services and components required to support mobile applications using cloud computing. It provides the reference architecture for mobile cloud [22].

Hyrax [23] is a platform derived from Hadoop that supports cloud computing on Android smartphones. Hyrax allows client applications to conveniently utilize data and execute computing jobs on networks of smart-phones and heterogeneous networks of phones and servers. Hyrax supports a Hadoop [24] cluster which is configured to run on Android phones. Running Hadoop on a cluster of phones is analogous to running Hadoop on a cluster of servers. It is a very interesting approach, but we this it is not suitable for Smart Cities applications, where data processing should be mode "centralized".

Mobile-Edge Computing (MEC) [25] provides cloud-computing capabilities at the edge of the mobile network. The main features are ultra-low latency, high bandwidth as well as real-time access to radio network information that can be leveraged by applications. Technically, it is a next step in the convergence of IT and telecommunications networking. As per ETSI vision, use cases include video analytics and Internet of Things. In other words, it is more than applicable for Smart Cities.

The key element of MEC is its application server, which is placed with the base station. This server provides computing capabilities, storage capacity, and connectivity. It looks like a localized cloud infrastructure. Via own Application Program Interfaces (API), this server provides access to the traffic and radio network information. This option lets application developers to tune their services. The above-mentioned API also allows real-time analytics. We think that MEC will be a mandatory part of 5G deployment and could be a part of Smart City technological stack. At this moment, it looks just as a promising technology from ETSI.

Cloudlet model [26] is similar to MEC. A cloudlet is a cloud datacenter that is located at the edge of the Internet. It is the same idea to support resource-intensive and interactive mobile applications by providing computing capabilities and storage capacity to mobile devices with lower latency.

As per its original idea [27], a cloudlet is a new architectural element that arises from the convergence of mobile computing and cloud computing. It represents the middle tier of a 3-tier hierarchy:

mobile device -> cloudlet -> cloud.

Some authors call it as a data center in a box with the main goal to bring the cloud closer [28]. The key features, usually listed in scientific papers are self-managing, good connectivity to the cloud, a logical proximity to the associated mobile devices, a usage of standard cloud technologies.

The logical proximity could be defined as low latency and high bandwidth. In terms of physical proximity it could be, for example, combined with Wi-Fi access point.

III. ON SELECTION CRITERIA

In this part, we would like to discuss the criteria for mobile cloud selection in Smart City projects. By our opinion, Smart Cities development adds own specific to this process.

Let us return back to the basic ideas behind MBaaS. Actually, they are very simple:

- mobile applications need a back-end

- back-end services are complex to build and test

- reuse the same back-end service can decrease time-to-market value for mobile applications.

So, the time to market is the key criteria here. Actually, the same is true for any API's pretended to the "standard". The standard API should simplify the development.

By the commercial reasons, Smart City core applications (Smart City platforms, Smart City SDK, etc.) in the most cases rely on open solutions and cannot use the commercial products without limitations.

In some countries (Russia is an example), collected data should be should be saved locally (cannot cross the borders). This requirement closes the usage Amazon and other popular public clouds. Technically, this restriction covers so-called personal data. But because many public measurements in Smart Cities are collected with smartphones, they fall into this category (smartphone's ID – IMEI, for example, links measurements to the owner).

The set of services for Smart Cities is not stable. We have to deal with the constantly updated list of services. Some old of them become irrelevant while new specifications require the newest development. Cities often involve non-professional developers from various areas into service (software) development. This once again emphasizes the requirement for simplicity and speed of development. Creating services should be cheap, as well as the abandonment of existing services should not be very expensive.

Most of the services for Smart Cities, at least, in the current vision, could be presented as on-demand Internet of Things (IoT) system. This IoT system contains several elements (e.g., sensors) at the edge, network function capabilities (middle tier), cloud services (backend). The typical situation for Smart Cities is a short but heavy workload. It is crucial to support data gathering in such cases and provide an end-to-end provisioning. And obviously, Smart Cities should avoid provisioning the above-mentioned elements separately and manually. A dynamic provisioning of resources for IoT systems requires so-called information-centric design [29]. An opposite approach is so-called host-centric architecture, which is tied to the actual host where particular functionality is available [30].

The information-centric approach leverages such functions as in-network caching, multiparty communication via replication, and (most important) interaction models where senders and receivers are decoupled. We can see several information-centric models. For example [31]:

- Data-Oriented Network Architecture (DONA) [32]
- Content-Centric Networking (CCN) [33],
- Publish-Subscribe Internet Routing Paradigm (PSIRP) [34]
- Design for the Future Internet and Scalable and Adaptive Internet Solutions [35]

In our paper, we present an experimental design for mobile cloudlets with a publish-subscribe model.

IV. MESSAGING CLOUDLETS

In this section, we describe our experimental design for information-centric architecture. This approach currently is testing with one mobile operator in Russia.

Our idea is to deploy publish-subscribe message system as cloudlet. As the particular example of high-throughput distributed messaging system, we choose Kafka system [36]. Kafka is a distributed message broker. Technically, message brokers are used for a variety of reasons. The two most important arguments in our case are: to decouple processing from data producers and to buffer unprocessed messages. We think these two features are very suitable for the short and heavy workloads, mentioned in Section III. The original use case for Kafka was to be able to rebuild a user activity tracking pipeline as a set of real-time publish-subscribe feeds. [37]. These feeds are available for subscription for a range of use cases including real-time processing, real-time monitoring, and loading into databases for offline processing and reporting. Actually, it is a perfect example of the measurements in Smart City applications.

In general, Kafka is a distributed publish-subscribe messaging system that is designed to be fast, scalable, and durable. At its core, Kafka maintains feeds of messages in topics. Kafka treats each topic partition as a log (an ordered set of messages). Producers (e.g., mobile applications) write data to topics and consumers (data proceedings) read from topics. Kafka is a distributed system, so, topics are partitioned and replicated across multiple nodes. The key abstraction in Kafka is a structured commit log of updates (Figure 5):

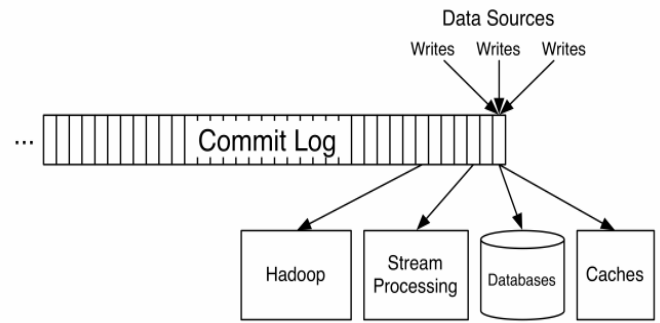

Figure 5. Kafka structure [38]

Messages in Kafka are simply byte arrays. It is possible to attach a key to each message, in which case the producer guarantees that all messages with the same key will arrive to the same partition.

Kafka does not attempt to track which messages were read by each consumer. Kafka retains all messages for a predefined amount of time, and consumers are responsible for tracking their location in each log. In our case, Smart City's data proceeding applications have got some predefined time to read measurements and react. By these principles, Kafka can support a large number of consumers and retain large amounts of data with very little overhead.

In our case, we are planning to use Kafka cluster (Figure 6). When communicating with a Kafka cluster, all messages are sent to the partition's leader. The leader is responsible for writing the message to its own in sync replica and, once that message has been committed, is responsible for propagating the message to additional replicas on different brokers.

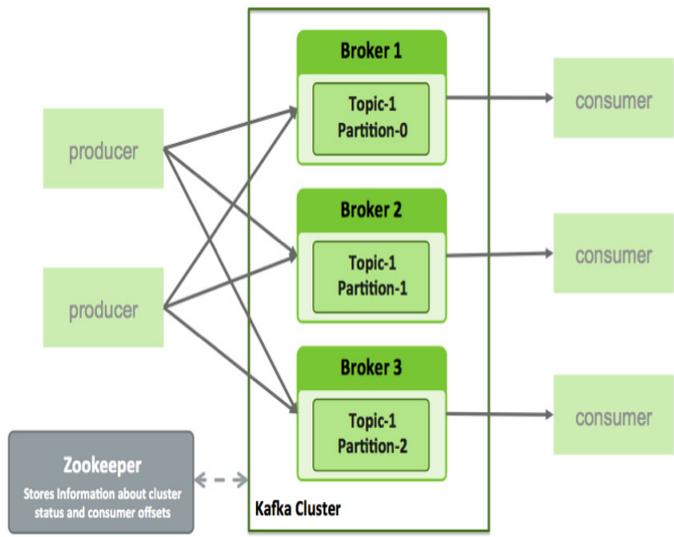

Figure 6. Kafka cluster [39]

The topic here corresponds to the particular service within Smart City, presented by the producer (e.g. mobile application).

Mobile clients (e.g., crowdsensing applications [40]) communicate with Kafka through REST proxy, based on Nginx. This proxy allows more flexibility for developers, and it significantly broadens the number of systems and languages that can access cloudlet (Kafka cluster).

As the backend data store, we propose Cassandra [41]. By our opinion, it is the best choice for time series data (typical representation for the measurements. And cloud-based Cassandra installation is a widely used choice [42].

## V. CONCLUSION

In this paper, we discuss mobile cloud development in deployment for Smart Cities application. We have discussed several opportunities as well selection criteria for cloud support in Smart City applications. We propose a design for mobile cloud solution based on the cloudlet approach. In our design, we use Kafka cluster as messaging based cloudlet, REST proxy for mobile clients and Cassandra as cloud data store. By our opinion, it has got advantages over existing proposals like FI-WARE due to its simplicity and usage of standard open source solutions in the background.

## ACKNOWLEDGMENT

We would like to thank people from Open Information Technologies Lab in Lomonosov Moscow State University for the valuable discussions.